# RCModel, a Risk Chain Model for Risk Reduction in AI Services

Takashi Matsumoto (Deloitte Touche Tohmatsu LLC/The University of Tokyo)

Arisa Ema (The University of Tokyo/RIKEN AIP Center)

6 July 2020

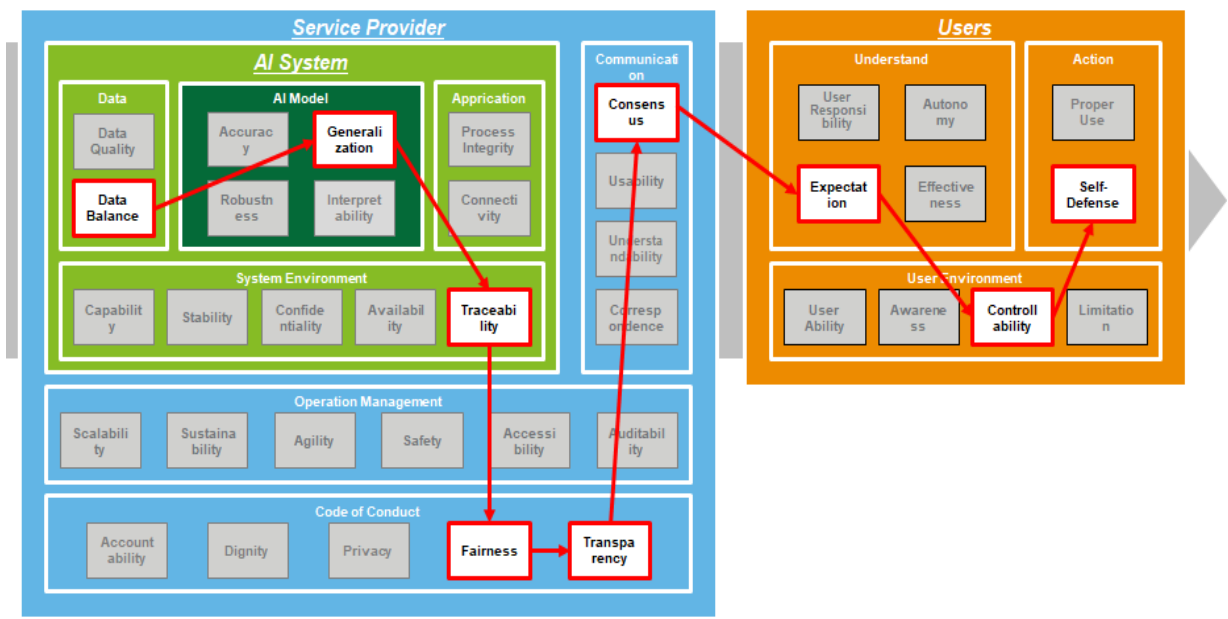

**Example of risk factors and risk chain**

**Executive Summary**

With the increasing use of artificial intelligence (AI) services and products in recent years, issues related to their trustworthiness have emerged and AI service providers need to be prepared for various risks. In this policy recommendation, we propose a risk chain model (RCModel) that supports AI service providers in proper risk assessment and control. We hope that RCModel will contribute to the realization of trustworthy AI services.

## Overview of RCModel

*1) Organization and structure of risk components*

There are a number of potential risk factors involved in provision of AI services. In RCModel, these factors are segregated into (1) technical components of the AI system, (2) components related to the code of conduct (including communication with users) of the service provider, and (3) components related to the user's understanding, behavior, and usage environment.

*2) Identification of risk scenarios and risk-contributing factors*

RCModel helps identify risk scenarios related to AI services, such as unfair decisions and uncontrollable accidents. It then identifies risk factors for priority risk scenarios.

*3) Visualization of risk chains and planning risk control*

Because it is difficult to reduce risk sufficiently on a factor basis, AI service providers can consider stepwise risk reduction by visualizing the relationships (risk chain) among the risk factors related to risk scenarios. This allows consideration of where a risk exists and its effective and efficient control.

## Policy Recommendations for Future Development and Implementation of AI Services Using RCModel

*Policy Recommendation 1: Enhance understanding of risk scenarios and factors*

Service providers need to properly understand the risk factors associated with their AI services. They should also pay attention to social incidents involving the use of AI technologies and recognize important risk scenarios.

*Policy Recommendation 2: Promotion of appropriate risk controls using RCModel*

AI service providers should formulate their risk control measures by analyzing RCModel's risk chain. It is neither necessary nor always possible to reduce all the risks identified; therefore, appropriate controls should be established within an enterprise based on factors such as magnitude of risks posed, technical difficulty, cost-effectiveness, and continuity.

*Policy Recommendation 3: Promoting and updating dialogue among stakeholders*

RCModel should be used to facilitate dialogue among AI service providers, AI developers, and users. In addition, a system should be established to clarify risk tolerance, create risk scenarios, structure risk factors, examine risk control models, and create a common understanding on the scope of each stakeholder's responsibility.

## 1. Issues and Aims

With the increasing use of artificial intelligence (AI) services[1] in recent years, issues related to their trustworthiness (e.g., unfair judgment, uncontrollable accidents, etc.) have emerged. In addition, AI service providers need to be prepared for a wide variety of risks, such as: decisions made by AI (especially deep learning) algorithms are not constant and can fluctuate; risk reduction when data provision or AI model development is outsourced and the AI service provider is not the only stakeholder; and the performance of an AI system may deteriorate or deteriorate due to misuse.

In this policy recommendation, guidelines related to trustworthy AI services are introduced in Chapter 2. In Chapter 3, a risk chain model (RCModel) is proposed as a model that enables AI service providers take appropriate risk control measures for their services. Chapter 4 presents a case study, Chapter 5 provides recommendations on how to use RCModel, and the final chapter outlines future issues and prospects.

## 2. Principles and Guides around Trustworthy AI

### 2-1. Overview of Principles and Practices

Previous research studies that systematically summarize the value provided by AI services include categorization of the principles presented by industry, academia, the public and private sectors[2], and arrangement of issues to consider ethical viewpoints from the development stage[3].

There have also been attempts to turn principles into practical guides[4]. For example, the European Commission[5] and the Dubai government[6] have provided self-assessment sheets on the risks posed by AI technologies. Singapore's Personal Data protection Commission (PDPC)[7], released the Model AI Governance Framework in January, 2020[8]. Their framework consists of internal governance structures and measures, determination of the level of human involvement in AI-augmented decision making, operations management, and stakeholder interaction and communication—each of which is explained with examples using actual companies (Master Card, GRAB, Facebook, etc.). In conjunction with the framework, self-assessment[4] and examples of corporate governance[9] are also published.

In Japan, the AI Network Society Promotion Council of the Ministry Internal Affairs and Communication

---

[1] In this policy recommendation, services using artificial intelligence (AI) will be the subject of discussion. In this context, "AI Services" refers to the provision of services that utilize the judgment of the AI model, including products such as AI speakers.

[2] A. Jobin, M. Ienca & E. Vayena: The Global Landscape of AI Ethics Guidelines, Nature Machine Intelligence, 1, 389-99, 2019.

[3] IEEE: Ethically Aligned Design First Edition, 2019.

[4] J. Morley, L. Floridi, L. Kinsey & A. Elhalal: From What to How: An Initial Review of Publicly Available AI Ethics Tools, Methods and Research to Translate Principles into Practices, Science and Engineering Ethics, 2019.

[5] High-Level Expert Group on AI (HLEG) of European Commission: Trustworthy AI Assessment List (Pilot Version), 2019. There are "Trustworthy AI Assessment List (pilot version)" with 129 items. https://ec.europa.eu/digital-single-market/en/news/ethics-guidelines-trustworthy-ai

[6] Smart Dubai: Ethical AI Toolkit, 2018. https://www.smartdubai.ae/initiatives/ai-principles-ethics

[7] Personal Data Protection Commission Singapore: Implementation and Self Assessment Guide for Organisations (ISAGO), 2020.

[8] PDPC: Model AI Governance Framework Second Edition, 2020.

[9] PCPC: Compendium of Use Cases: Practical Illustrations of the Model AI Governance Framework, 2020.

had released AI utilization guidelines[10] in August 2019. This document defines the 10 principles of proper utilization, data quality, collaboration, safety, security, privacy, human dignity and individual autonomy, fairness, transparency, and accountability, and it also defines the matters that each AI service provider/business user/data provider/consumer user should keep in mind.

## 2-2. Challenges in Translating Principles into Practices

These principles and guidelines provide a broad understanding of the general issues surrounding trustworthy AI services. However, because the risks to be considered for each AI service vary, using principles and guidelines as a simple checklist may diminish focus on important risks. In addition, it is considered that operational costs will grow so that AI services are not being implemented effectively.

Moreover, a risk related to a trustworthy AI service can be actualized not only by the AI model but also by multiple other factors such as learning data, input/output of data, users, and usage environments (e.g., in the case of AI fairness, data biases, algorithms bias, and biases in user can be risk factors). The risks are related, and sometimes there can be a trade-off between them. Therefore, it is desirable to comprehensively consider multiple factors in risk management.

If all processes such as data acquisition, AI model development, and service delivery have been conducted by one company like Google and Facebook, companies can take an optimal approach to all risk factors. However, in Japan, there are many cases in which service providers, developers of AI models, providers of execution environments, and data providers are different. Therefore, in order for AI service providers to comprehensively examine all risk factors, a framework for dialogue among relevant stakeholders is necessary.

In this policy recommendation, we provide a model that enables AI service providers to visualize the relationship between risk factors related to the trustworthy AI services and discuss optimal risk responses through dialogue with relevant stakeholders (e.g., AI system developers and users).

## 2-3. Risk Assessment and Control

As AI services are deployed in various fields, to ensure their trustworthiness, it is first necessary to sort out the types and magnitudes of risks that can occur with the consumption of the services as well as who will incur these risks[11]. After the type of risk and the stakeholders concerned are examined, the next step is to identify important risk scenarios by taking into account the degree of impact and the probability of occurrence, and then comprehensively understand the risk factors related to the identified risk scenarios (i.e., risk assessment). The relationship between the risk factors is then analyzed and an effective and efficient risk response is implemented (i.e., risk control).

A single risk response alone may not be sufficient; therefore, layering multiple risk responses is needed. For example, "multilayered defense," such as the Swiss cheese model, are often used against cyber-attacks[12].

---

[10]  MIC, The Conference toward AI Network Society: Overview of 2019 Report, 2019.
[11]  In areas such as the financial sector, where risks are actively undertaken, a risk appetite framework (RAF) that expresses the type and amount of risks to be willingly accepted as "risk appetite" has been constructed in order to achieve the purpose of the organization. For reference, "Establishment of a Risk Appetite Framework" (T. Oyama, 2015, Chuokeizaisha).
[12]  J. Reason, E. Hollnagel & J. Paries: Revisiting the «Swiss Cheese» Model of Accidents, 2006.

The multilayered defense strategy of the U.S. National Security Agency (NSA) advocates the realization of risk countermeasures in each layer of "People element," "Technology element," and "Operations element[13]."

In addition, the Internal Control Reporting System (Japanese Sarbanes-Oxley Act: J-SOX)[14] is a useful tool for companies to implement risk control. J-SOX reviews risk scenarios in key processes related to financial reporting and seeks to reduce risk. For example, when the risk scenario of "accounting for fictitious sales" is considered in the "sales process," risk control is performed in multiple operations: (1) sales order received: credit confirmation, (2) sales delivery: existence confirmation, (3) sales and billing: inspection by customers, and (4) collection of claims: receivable balance confirmation. As each control does not necessarily reduce the risk sufficiently by itself, multiple controls (four stages in an "accounting for fictitious sales" risk scenario) are introduced to reduce the risk. From this viewpoint, this model also applies stepwise risk reduction to the model.

## 3. Overview of the Risk Chain Model

As described above, risk management related to the trustworthiness of AI services requires comprehensive consideration of multiple risk factors. In the risk chain model (RCModel), the relationship among each risk factor in the AI service is visualized using a chain.

In order to utilize RCModel, the following framework is examined. Identify the factors of a structured RCModel (3-1); identify risk scenarios for individual AI services and prioritize the scenarios to be addressed (3-2); next, identify the risk factors that constitute each risk scenario (3-3); then, analyze the relationships among these risk factors (3-4); finally, implement appropriate controls against each risk factor (3-5). Each of these steps is discussed below.

### 3-1. Structuring Risk Factors

There are a number of keywords related to risk reduction and each guideline/set of principles defines them slightly differently. Some studies even categorize them, building on previous guidelines[15].

However, even if the same expressions are used to explain keywords such as fairness and privacy, the scope and content of each guideline may differ. For example, the Japanese Government's AI Utilization Guideline[10] includes "traceability" and "explainability" as a subcategory" under "transparency," whereas the European Commission's Ethics Guideline for Trustworthy AI[5] adds "communication" to "traceability" and "explainability," and Singapore's Model AI Governance Framework[8] treats "explainability" and "transparency" as separate categories. As described above, because the expressions used in each guideline are different, it is difficult for the AI service provider to determine which guideline to refer.

Further, for successful risk measurement, it is desirable that the keywords are classified in a way that it is possible to grasp whether it attributable to AI systems (technology), AI service providers (operation), or users.

---

For example, in order to realize "explainability," the responses are different depending on "interpretability" for visualizing the judgment basis of the AI model and "understandability" for making the expression understandable by a person depending on the AI service provider.

Therefore, we organized the principles and keywords mentioned in AI ethics and governance guidelines published in Japan and abroad, and created a model by classifying the risk factors for AI services into three layers: (1) AI systems, (2) AI service providers, and (3) users (Figure 1). The first layer (AI systems) is the technical layer and includes components for AI models, data, rule-based applications, and system environments. The second layer (AI service providers) is the layer of service operation for the user that includes the AI systems as well as the code of conduct, operations, and communication. The third layer (users) represents the direct users of the AI service and includes user understanding, action, and user environment. See Appendix 1 for a list of guidelines used to organize risk factors and see Appendix 2 for definitions of each component and risk factor.

However, social incidents and the development of new technologies may change the interpretation of these risk factors and new risks may emerge. Also, in some cases, the arrangement of the risk factors in the model may change. In this report, the factors are arranged in each layer based on current considerations.

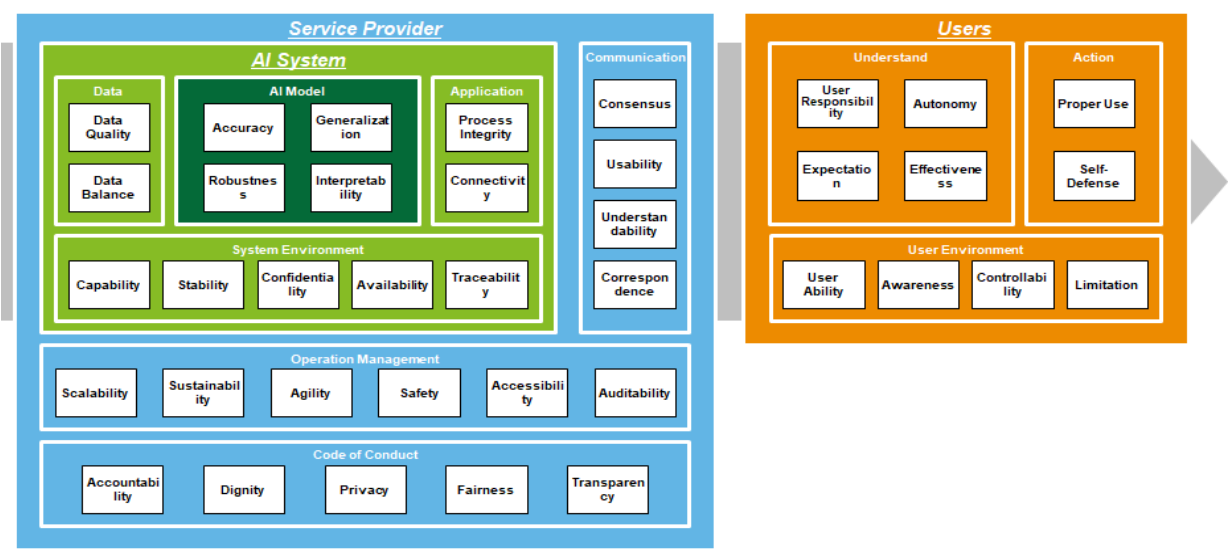

**Fig 1. RCModel factors and structure**

### 3-2. Risk Assessment and Creating Scenarios

As risks related to the trustworthiness of AI services include not only the quality but also ethical and legal issues, it is difficult to generalize risk scenarios and create checklists. Therefore, when considering risk scenarios, it is necessary for people involved in AI services to develop scenarios from various perspectives. In creating risk scenarios, the following steps could be used: (1) investigate the general risk perspectives indicated in each guideline (Appendix 1), (2) refer to incidents and accidents that could undermine the trustworthiness of the AI service, and (3) form a group consisting of people from various industries and attributes, such as technology, legal affairs, and sales to discuss risk scenarios. In Japan, some companies have set up committees of experts outside the company to discuss the ethical, legal, and social implications of AI services[16].

Table 1 provides examples of factors that could undermine the trustworthiness of an AI service, including (1) the concerns expressed by various previous reports and guidelines and (2) specific incidents and cases.

**Table 1. General Case of Risks around AI Services**

| Factor | (1) Concerns expressed in guidelines and reports | (2) Specific incidents and cases |
|---|---|---|
| **Fairness** | The risk of making unreasonably negative judgments about user groups with certain attributes[17] | Extremely negative judgments were made against women in AI recruitment[18] |
| **Robustness** | The risk of making extremely erroneous judgments due to the presence of minute noise in the data, causing disadvantage or damage to users[19] | It has been pointed out that when a self-driving car recognizes a road sign, there is a risk of it making a very bad decision because of minute noise that cannot be judged by human eyes[20] |
| **Explainability** | The risk of failure to explain the reasoning behind AI decisions will result in insufficient accountability in the event of any conflict[21] | In the use of AI in medicine, concerns have been raised over whether physicians will be able to correctly interpret the judgment of AI as a "black box" and adequately explain the result to patients[22] |
| **Proper use** | The risk that the performance of the AI service will deteriorate due to inappropriate use by the user, and the AI service will be disadvantageous to another user[23] | In a conversation with a chatbot, a specific user made many discriminatory remarks, and the chatbot itself made many controversial remarks[24] |

---

[16] As the purpose of this policy recommendation is to propose a framework for the risk chain model, which will be discussed later, consideration of risk scenario derivation methods, assessment methods, and corporate governance is not included in the paper. However, even in Japan, there are committees that assess the risks of AI services, for example, Fujitsu established the External Advisory Committee on AI Ethics and ABEJA has created the Ethical Approach to AI (EEA).

[17] MIC: AI Utilization Guidelines[7] "8) Principle of fairness"

[18] Amazon Reportedly Killed an AI Recruitment System Because It Couldn't Stop the Tool from Discriminating Against Women, Fortune. https://fortune.com/2018/10/10/amazon-ai-recruitment-bias-women-sexist/

[19] I. J. Goodfellow, J. Shlens & C. Szegedy: Explaining and Harnessing Adversarial Examples, 2014. https://arxiv.org/abs/1412.6572

[20] I. Evtimov, K. Eykholt, E. Fernandes, T. Kohno, B. Li, A. Prakash, A. Rahmati & D. Song: Robust Physical-World Attacks on Machine Learning Models, 2017. https://arxiv.org/abs/1707.08945v3

[21] MIC: AI Utilization Guidelines[7] "10) Principle of accountability"

[22] Japan Medical Association: The 9th Science Promotion Committee's report "AI and Medicine," 2018. http://dl.med.or.jp/dl-med/teireikaiken/20180620_3.pdf

[23] MIC: AI Utilization Guidelines[7] "1) Principle of proper utilization"

[24] Microsoft Takes Chatbot Offline after It Starts Tweeting Racist Messages, Time, 2016

From this perspective, risk scenarios for each AI service could be identified; however, it is difficult to address all risk scenarios. Therefore, risk scenarios should be prioritized based on their potential impact and likelihood[25].

**3-3. Identification of Risk Factors**

For the identified risk scenario (3-2), the risk factors need to be identified (3-1) and the relationship between them is visualized through a risk chain (3-4). By visualizing the risk chain, AI service providers can determine which risk factors can be controlled. In order to link risk scenarios to each risk factor, the factors attributable to risk scenarios are decomposed and relevant components are selected based on the definitions in Appendix 2.

**3-4. Visualization of Risk Chains**

Next, the risk factors identified for a risk scenario are connected to each layer (AI systems, AI service providers, and users) based on the order in which the risk factors become apparent. The risk chain is not necessarily unidirectional and single-line, and the AI system is not necessarily the starting point. It is important to deepen the discussion on the risks related to AI services by accumulating and brushing up on the discussions among the stakeholders concerned based on the risk chain as a common base.

**3-5. Examination of Control Using Risk Chain**

By visualizing the risk chain, AI service providers can correlate each factor and consider effective and efficient risk mitigation/control measures. It would be possible to further reduce the risk if countermeasures could be implemented for all factors. However, in practice, it is assumed that some controls do not function effectively, are too costly, or cannot be implemented due to constraints such as data on specific communities being difficult to obtain, the AI model is highly complex, or in sufficient judgment skills of users. As a result, AI service providers do not have to consider a control against every risk factor in the risk chain. Instead, priority should be given to controls that are highly effective in reducing risk and that are easy to manage and are cost-effective.

In addition, the risk factors and controls may be different, even if the same risk chain is drawn, due to difference in service structure, scope of responsibility, control, and technical abilitydifficulty of the AI service provider. Therefore, it is important to plan effective controls based on needs, abilities, and concerns of all stakeholders.

**4. Case Study: RCModel Usage in Hiring AI Services**

This section presents an example of the use of RCModel by utilizing a specific case, which invo lves a company's personnel department using an AI service to determine which applicants to hire fr om an entry sheet.

---

**4-1. Specific Information**

   This case is not intended for any particular AI service but is a hypothetical example of how to use RCModel. However, in order to give reality to scenario creation, we created the following situation as an example.

Overview of the case

- This is an AI service used as reference information when judging the selection of documents for the entry sheet for the human resource recruitment department in a global company of *Company A*.

- The AI development department of *Company A* of the AI service provider receives past entry sheet data and the result of pass/fail judgment from a personnel department of *Company A* (including overseas group companies), which is a business user. They created a learning model for judging pass/fail using machine learning (classification model).

- It is evaluated based on "precision" (the percentage of successful applicants who have passed an interview and have been offered a job offer), and 70% is set as the expected value. The "precision" is used as an evaluation index because the "recall" (the percentage of people who didn't get a job offer because they didn't pass the AI screening process) does not provide enough data for the examination.

- The personnel department of *Company A* reads the entry sheet (by electronic file) of the applicant (both new graduates and mid-career graduates) into the learning model and can confirm the judgment result (pass/fail judgment) of AI on its own personal computer (via a browser). It is to be noted that not only the pass/fail judgment but also the keyword affecting the judgment is highlighted and displayed in the entry sheet on the output screen. The person in charge of the personnel department sets up a pass/fail judgment using the judgment of the AI as reference information, obtains the approval of the head of the personnel department, and notifies the applicant of the pass/fail.

- The stored real data are appropriately added as learning data at the time of input of determination result, and the learning model is updated and deployed daily. However, the AI model in the production environment is not automatically updated when the correct answer rate is less than 70% as a result of cross-validation testing in the learning process. The AI model stores versions from the past year.

**Table 2. Data input in the AI service**

| Data | Purpose | Collection Method | Data Manager | Including Privacy Data |
|---|---|---|---|---|
| **Past entry sheet data** | Learning | Entry sheet data submitted by the applicant to *Company A* personnel department and pass/fail label | Head of personnel department, *Company A* (*Company A* private cloud environment) | Yes (including sensitive personal information) |
| **Newest entry sheet data** | Production | Entry sheet data submitted by the applicant to *Company A*'s group personnel department | Head of personnel department, *Company A* (*Company A* private cloud environment) | Yes (including sensitive personal information) |

**Table 3. Output of the AI service**

| Users | Person in charge of personnel department at *Company A* |
|---|---|
| Output | Pass/Fail |
| Output method | When the entry sheet of the applicant is input on the terminal of the person in charge of the personnel department of *Company A*, the judgment of the document selection is output |
| Expected accuracy | Precision: 70%<br>*The percentage of those who actually received job offers among those who passed the screening process |
| User judgment | Yes |
| Output of evidence information | Keywords that have had a strong impact on the decision are highlighted in the entry sheet |
| Safety risk | No |
| Connection with external system | No |

**Fig 2. Relationship between the AI service and users**

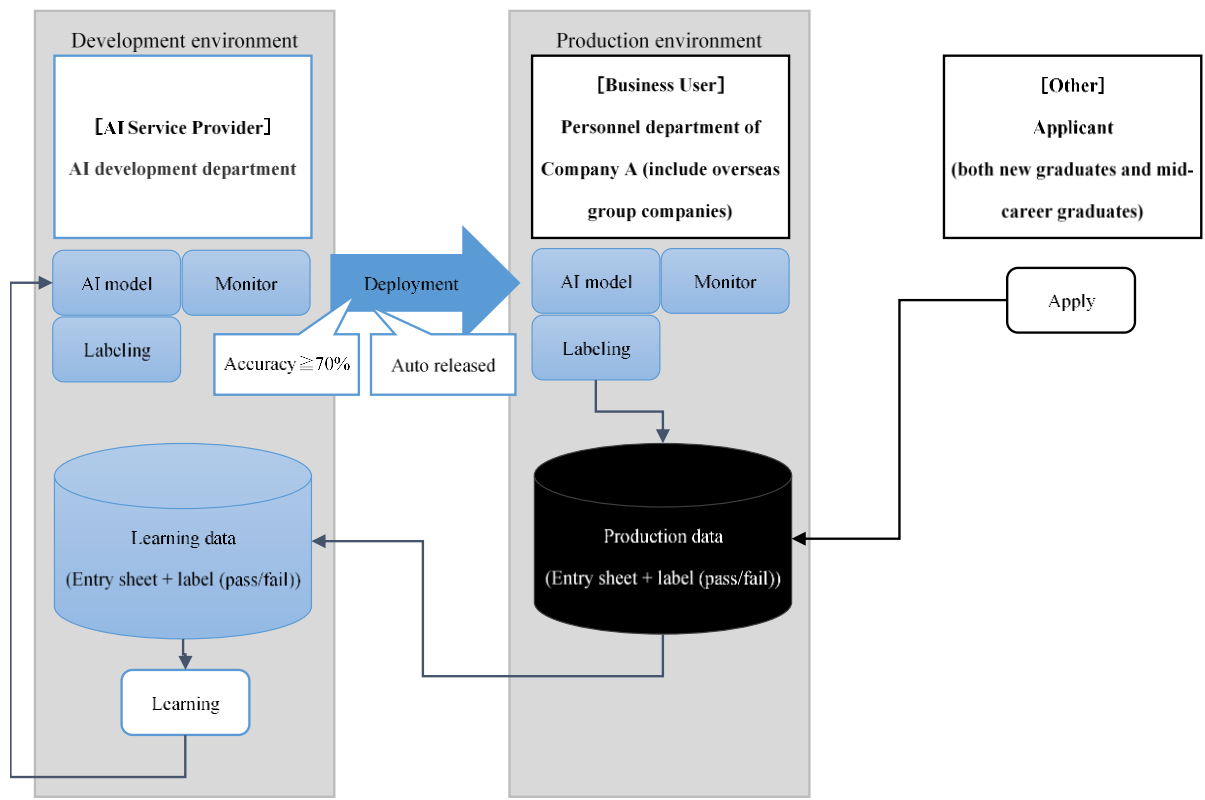

## 4-2. Risk Assessment and Scenario Creation

First, we identified the risk scenarios that apply to this case, from a general point-of-view, in accordance with MIC's Utilization Guideline of AI[10]. In addition, we referred to the reports and recommendations on the issues related to hiring AI[26]. Table 4 shows the results of examining "degree of influence" and "probability of occurrence" for each risk scenario and arranging them in order of priority.

**Table 4. Possible Risk Scenarios**

| No | Risk Scenario | References |
|---|---|---|
| R001 | Produces incorrect predictions for a specific country/region/race/gender/age | MIC: AI Utilization Guidelines[7] "8) Principle of fairness" Draft Recommendation on Profiling Governance [24] |
| R002 | Recruiters rely too much on AI decisions, not noticing AI mistakes and making inappropriate final decisions | MIC: AI Utilization Guidelines[7] "7) Principle of human dignity and individual autonomy" |
| R003 | A person in charge of recruitment uses AI services many times to identify entry sheets and key phrases for which the AI service makes a pass judgment with high probability, and the person illegally sells them outside the company to human resource agents, etc. | MIC: AI Utilization Guidelines[7] "1) Principle of proper utilization" |
| R004 | Inaccurate personnel department people's feedback to AI (anonymization of personal information in entry sheet and label setting of pass/fail) causes AI to make inappropriate decisions | AI Utilization Guidelines[7] "1) Principle of proper utilization" |
| R005 | Because hiring policies and personalities of applicants (ex. race) vary by company in the region, appropriate forecasts cannot be made without the preparation of training data at each group company | Raised by case evaluator (distribution of appropriate learning data) |
| R006 | A slight difference in the character information used in the entry sheet (ex. differences in punctuation) can significantly change the AI's decision | Raised by case evaluator (Robustness Risk[18]) |
| R007 | When privacy information is mishandled and leaked, appropriate measures cannot be taken, resulting in increased damage and legal violations (violations of the Personal Information Protection Act) | MIC: AI Utilization Guidelines[7] "6) Principle of privacy" |

---

[26] Although HR-Tech is not introduced in these policy recommendations, the "Draft Recommendation on Profiling Governance" published by the Personal data + Alpha Research Group in February 2019 (in Japanese), addresses concerns and examples related to recruitment profiling (https://www.shojihomu-portal.jp/nbl20190222). Upturn published a report in December 2018 on equality and bias in hiring AI. M. Bogen and A. Rieke, Help Wanted: An Examination of Hiring Algorithms, Equity, and Bias, 2018.

Because these risk scenarios are derived for use in the technologies and applications described in 4-1, the risk scenarios and priorities derived for the same AI may differ if the usage and service provision methods change.

**4-3. Identify Risk Factors for Each Risk Scenario**

After identifying and prioritizing the risk scenarios, we need to identify the risk factors associated with each risk scenario. For example, in this case, we identified risk factors for the risk scenario "R001 Produces inappropriate predictive results for a specific country/region/race/gender/age." To take into account the order in which the risks appear when they are linked in the risk chain, risk factors are identified in three stages: (1) prevention: factors that do not prevent risk, (2) discovery: factors that fail to discover the realization of risk, and (3) response: factors that are not adequately addressed when risks are identified" (Table 5).

**Table 5. Risk Factors at Each Stage**

| Prevention | Detection | Response |
|---|---|---|
| ■ [AI system] Data bias prevents fair judgment (data balance) <br><br> ■ [AI system] Generalization performance of the algorithm is impaired and fair judgment is not made (generalization) | ■ [AI system] AI judgment basis unknown (interpretability) <br><br> ■ [AI system] Inability to validate AI decisions (traceability) <br><br> ■ [Service provider] Points to consider when making fair judgments are not clear, and judgment scales vary greatly from person to person (fairness) <br><br> ■ [Service provider] Inability to visualize negative judgments when they tend to be negative for a particular group (transparency) <br><br> ■ [Service provider] The user does not recognize the points to be noted, and the user does not make a fair decision (consensus) | ■ [User] The user does not recognize the points to be noted, and the user does not make a fair decision (consensus) <br><br> ■ [User] Fair judgment is not made unless a person is aware of making a negative judgment on a specific group (expectation) <br><br> ■ [User] Discriminatory judgment is not corrected because the decision-making process is unclear (controllability) <br><br> ■ [User] Fair selection is not made in the final judgment （proper use） |

For the (1) prevention stage, two risk factors are identified for the AI system layer: "data bias prevents fair judgment (data balance)" and "algorithmic bias prevents fair decisions (generalization)."

For the (2) discovery stage, two risk factors are identified for the AI system layer: "don't know why AI is judged (interpretability)" and "inability to validate AI decisions (traceability)." For the AI service provider layer, "points to consider when making fair judgments are not clear, and judgment scales vary greatly from person to person (fairness);" "inability to visualize information when there is a tendency to make negative decisions for a particular group (transparency);" and "to solve the problem that fair judgment is not performed on a user side by not performing mutual recognition of attentions on the user side (consensus)" are identified.

For the risk factors in the response stage, in the user (person in charge of recruiting) layer, "undue reliance on AI decisions (human autonomy)" and "fair judgment is not possible unless a person is aware of making a negative judgment on a specific group (expectation)" are identified as matters to be "understood" by the user. Also, "indistinct decision-making process does not correct discriminatory decisions (controllability)" is identified as the "user environment," and "fair selection is not made at the final judgment stage (proper use)" is identified as the "action" of the user.

### 4-4. Visualization of Risk Chains

The factors associated with the risk scenarios and the identified items are connected in Figure 3. According to the order of the factors identified in the previous step (4-3): "Data Balance" → "Generalization" → "Interpretability" → "Traceability" in the system environment appear in the first layer, AI system. In the second layer, AI service provider, the following chain is created: "Fairness" → disclosure of necessary information "Transparency" → "Consensus" with the user. Finally, in the third layer: the user is connected to the "action" of the user as " Human Autonomy" → "Expectation" → "Controllability" → "Proper Use."

The risk chain draws a line in a single direction, starting from the AI system, based on the order of transmission of information, but the line does not have to be unidirectional and single. And it does not have to start from AI system either.

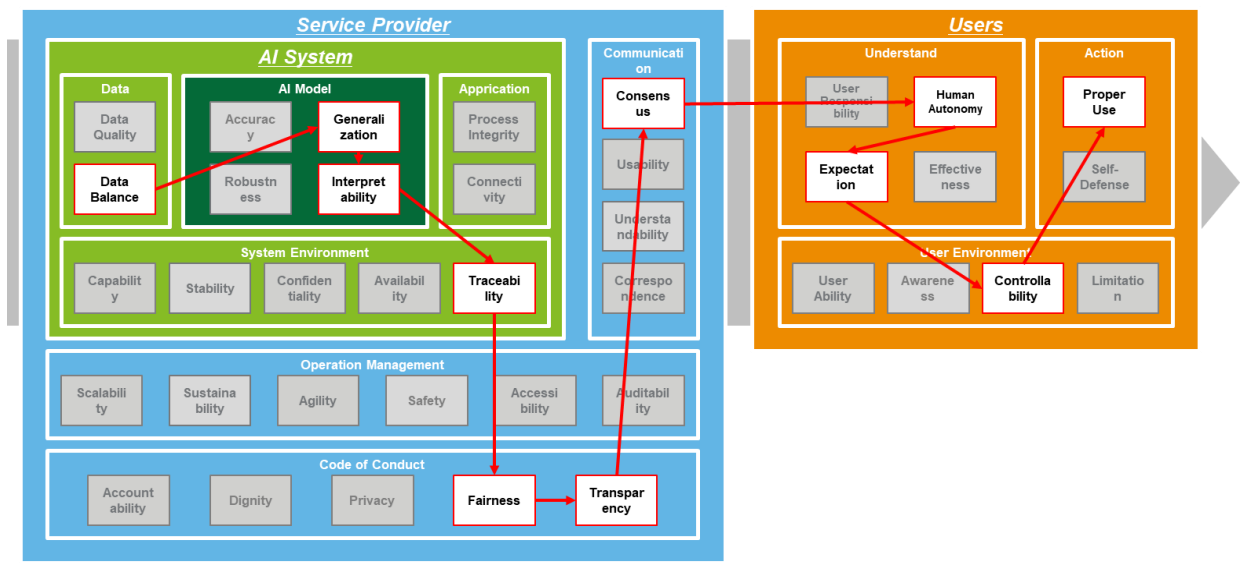

**Fig 3. Example risk chain for the case study**

**4-5. Examination of Risk Controls Using Risk Chain**

Table 6 summarizes the control proposals identified for each risk factor in each of the three layers. In deriving these controls, the previous reports and guidelines referenced in the risk scenario development and responses to social incidents are also helpful.

In addition, RCModel is not used to implement all control measures but is used to examine the controls that are highly effective in reducing risk and are easy to operate among the measures listed here. It can also be used as a basis for identifying controls that can be implemented by stakeholders and explaining them to AI system developers and users.

**Table 6. Example of Factors Controls in a Risk Scenario of the Case Study**

| Layer | Factor | Control |
|---|---|---|
| AI systems | Data balance | ● Adjust the ratio of data (ex. ratio of male to female) so that the learning data is not biased to a specific layer |
| | Generalization | ● Excluding feature quantities (ex. gender, nationality) that lead to unfair judgments from explanatory variables of the AI model |
| | Interpretability | ● Enable output of model basis (ex. importance of feature quantity) information |
| | Traceability | ● Preserve fairness information (ex. comparison of judgment results by attribute) during model learning<br>● Preserve information on grounds for judgment at the time of utilization (ex. features that strongly influence judgment results) |
| Service providers | Fairness | ● Compile points of concern for fairness (ex. gender, nationality) concerning applicants and make them known to the stakeholders |
| | Transparency | ● Disclose information that users (ex. personnel department staff) should be aware of, such as a tendency to make negative judgments on specific groups |
| | Consensus | ● Agree with users on the prediction accuracy of AI services, precautions (ex. making negative judgments on specific groups), and user responsibilities (ex. final judgment) |
| Users | Human autonomy | ● Based on the considerations provided by the AI service, make a decision not to rely excessively on AI |
| | Effectiveness | ● Confirm prediction accuracy and user considerations |
| | Controllability | ● Clarify which users have the authority to make final pass/fail decisions<br>● Representative of personnel department decide to accept AI's judgment as the final judgment |
| | Proper use | ● Confirm whether or not the judgment discriminates against a specific group, and make a final judgment on whether it is acceptable |

## 5. Policy Recommendations

Applying Chapter 3's overview, Chapter 4 used hypothetical examples to illustrate the use of RCModel for a practical use case. RCModel enables AI service providers to plan their risk response (controls). Recommendations for future development and implementation of AI services using RCModel are as follows:

*Policy Recommendation 1: Enhance understanding of risk scenarios and factors*

Service providers need to properly understand the risk factors associated with their AI services. They should also pay attention to social incidents involving the use of AI technologies and recognize important risk scenarios.

*Policy Recommendation 2: Promotion of appropriate risk control using RCModel*

AI service providers should formulate their risk control measures by analyzing RCModel's risk chain. It is neither necessary nor always possible to reduce all the risks identified; therefore, appropriate controls should be established within an enterprise based on factors such as magnitude of risks posed, technical difficulty, cost-effectiveness, and continuity.

*Policy Recommendation 3: Promoting and updating dialogue among stakeholders*

RCModel should be used to facilitate dialogue among AI service providers, AI developers, and users. In addition, a system should be established to clarify risk tolerance, create risk scenarios, structure risk factors, examine risk control models, and create common understanding on the scope of each stakeholder's responsibility.

## 6. Future Works and Developments

The policy recommendations provide a model for AI service providers to explore, interact with, and explain optimal risk control using risk chains among stakeholders. Also, the usage of the model was explained using a case study.

RCModel provides a framework to structure risk factors (3-1), examine risk scenarios (3-2), identify drivers for each risk scenario (3-3), visualize risk chains (3-4), and examine controls (3-5). It is important to carry out risk assessment and control with various stakeholders in this order from the viewpoint of facilitating issues. In addition, even if there is a minor change in the technology or mechanism, by returning to this framework, it is possible to visualize what should be changed in terms of scenarios and the way of connecting chains, which is effective from the viewpoint of transparency of discussion.

The policy recommendations provided a framework for risk assessment and control as a basis for discussion, but the methodology for discussion and methodology such as scenario derivation, chain connection, and control consideration will be systematized through further case studies. In addition, it is necessary to periodically review the risk factors in light of technological progress and social expectations. It is also important for society as a whole to discuss what kind of risks are acceptable and to what extent, especially in AI services used fields such as medical care and transportation. These issues should be discussed through risk assessment and control feedback, and it will be a future task to develop not only communication among stakeholders but also communication methods for the society.

The policy recommendations only focused on the risk control of AI services and products so far. However, AI service providers are required to be resilient to risks that are not assumed and to be able to respond to ongoing changes, and governance at the organizational level should also be considered. Examples include the formulation of policies, the securing of human resources and systems with expertise, the construction of development and operation processes, and the establishment of risk appetite. Furthermore, regarding AI services in high-risk areas such as transportation, social infrastructure, medical care, and public institutions, it is necessary to consider a third-party evaluation approach for AI services in order to objectively verify their reliability[27].

In order for AI services to be trusted and disseminated to society, governance in the stages of AI development, provision of services, and communication with users is essential, and we hope that RCModel will be used as a tool for prior and subsequent risk assessment and as a dialogue tool among stakeholders.

**Acknowledgments**

In writing this paper, we received advice from the Technology Governance Unit, Institute for Future Initiatives, the University of Tokyo, and from Deloitte Touche Tohmatsu LLC. We also received invaluable opinions from members of the Japan Deep Learning Association Public Affairs Committee. This study was supported by JST-RISTEX (Grant Number JPMJRX16H2) and Toyota Foundation (Grant Number D18-ST-0008).

---

[27] The 2019 report by AI Network Society Promotion Council of the Ministry Internal Affairs and Communication.

**Appendix 1: Reference List**

Guidelines referred for AI ethics and governance (*Published by: Name of Reference Material, Country/organization, date of publication*):

**Appendix 2: Layers, Components, and Factors**

| Layer | Component | Factor | Description |
|---|---|---|---|
| AI system | AI model | Accuracy | Predictive performance |
| | | Generalization | Generalization performance (impact of algorithmic bias) |
| | | Robustness | Noise resistance |
| | | Interpretability | Model interpretability |
| | Data | Data quality | Data integrity and timeliness |
| | | Data balance | Impact of data bias |
| | Application | Process integrity | Application integrity of rule-based logic |
| | | Connectivity | Protocol for connection with external systems |
| | System environment | Capability | Processing performance or system scalability |
| | | Stability | Stable running with error collection and reproductivity |
| | | Confidentiality | System confidentiality |
| | | Availability | System availability |
| | | Traceability | Transaction traceability or detection errors |
| Service provider | Code of Conduct | Accountability | Accountability of service providing |
| | | Dignity | Protection of the rights of user decision |
| | | Privacy | Protection of privacy |
| | | Fairness | Non discrimination |
| | | Transparency | Appropriate information visualization |
| | Operation | Scalability | Service scalability |
| | | Sustainability | Maintain the quality of service |
| | | Agility | Agile process for development |
| | | Safety | Harmless |
| | | Accessibility | Access control and authentication |
| | | Auditability | Internal and external auditability |
| | Communication | Consensus | Consensus between service provider and users |
| | | Usability | Easy to use |
| | | Understandability | Easy to understand |
| | | Correspondence | Cooperation with user and stakeholders including external specialist |
| Users | Understand | User responsibility | User responsibility |
| | | Expectation | Expectation of the performance of AI service |
| | | Human autonomy | Human autonomy |
| | | Effectiveness | Effectiveness of the risk from AI service |

|  | Action | Proper use | Proper use |
|---|---|---|---|
|  |  | Self-defense | Self-defense |
|  | User environment | User ability | Literacy, experience, and skill |
|  |  | Awareness | Recognize the AI existence |
|  |  | Controllability | User options to control |
|  |  | Limitation | Restrict user's wrong option |